\documentclass[11pt]{cernrep}
\usepackage{graphicx}
\usepackage{here}

\begin{document}

\title{On the Signal Significance in the Presence of Systematic and 
Statistical Uncertainties}
\author{S.I. Bityukov}
\institute{Institute for High Energy Physics, Protvino, Moscow Region, 142281, Russia}
\maketitle
\begin{abstract}
{The incorporation of uncertainties to calculations of signal significance 
in planned experiments is an actual task. We present a procedure of taking 
into account the effects of one sided systematic errors 
related to nonexact knowledge of signal and background cross sections
on the discovery potential of an experiments. 
A method of a treatment of statistical errors of
the expected signal and background rates is proposed. The interrelation 
between Gamma- and Poisson distributions is demonstrated.}
\end{abstract}

\section{Introduction}

One of the common goals in the forthcoming experiments is the search for new 
phenomena. In estimation of the discovery potential of the planned experiments
the background cross section (for example, the Standard Model cross section)
is calculated and, for the given integrated luminosity $L$, the average 
number of background events is $n_b = \sigma_b \cdot L$.
Suppose the existence of new physics 
leads to additional nonzero signal cross section $\sigma_s$ with the same 
signature as for the background cross section  that 
results in  the prediction of the additional average number of signal events
$n_s = \sigma_s \cdot L$ for the integrated luminosity $L$.
The total average number of the events is 
$<n> = n_s + n_b = (\sigma_s + \sigma_b) \cdot L$.
So, as a result of new physics existence, we expect an excess 
of the average number of events. The probability of the realization 
of $n$ events in the experiment is described 
by Poisson distribution~\cite{PDG,Eadie}

\begin{equation}
f(n; \lambda)  = \frac{{\lambda}^n}{n!} e^{-\lambda}.
\end{equation}

\noindent

In the report the approach to determination of the ``significance'' 
of predicted signal on new physics in concern to the predicted background
is considered. This approach is
based on the analysis of uncertainty~\cite{Bit1, Bit2}, which will take 
place under the future hypotheses testing about the existence of
a new phenomenon in Nature. 
We consider a simple statistical hypothesis $H_0$: {\it new physics is present 
in Nature} (i.e. $\lambda = n_s + n_b$) against a simple alternative hypothesis
$H_1$: {\it new physics is absent} ($\lambda = n_b$). 
The value of uncertainty is defined by the values of
the probability to reject the hypothesis $H_0$ when it is true 
(Type I error $\alpha$) and the probability to accept the hypothesis $H_0$ 
when the hypothesis $H_1$ is true (Type II error $\beta$). 
The concept of the ``statistical 
significance'' of {\it an observation} is reviewed in the ref.~\cite{Sinervo}.
All considerations in the paper are restricted to the most simple case 
of one channel counting experiment. More advanced statistical analysis 
based on other technique can be found, for example, in the refs.~\cite{Junk}.

\section{``Signal significance'' in planned experiment}

``Common practice is to express the significance of an enhancement by 
quoting the number of standard deviations''~\cite{Frodesen}.
Let us  define the ``signal significance'' 
(see, for example, ref.~\cite{Narsky}) as ``effective significance'' 
$s$~\cite{Bit0} 

\begin{equation}
\sum_{n = n_0+1}^{\infty}f(n; n_b) =
\frac{1}{\sqrt{2\pi}}\int_{s}^{\infty}exp(-x^2/2)dx ,
\end{equation}   
where $n_0$ is the critical value for hypotheses testing 
(if the observed value $n \le n_0$ then we reject $H_0$ 
else we accept $H_0$). In this case the system 

\begin{equation}
\beta =  \sum^{\infty}_{n=n_0+1} f(n; n_b) \leq \Delta
\end{equation}

\begin{equation}
1 - \alpha = \sum ^{\infty}_{n = n_0 + 1}
f(n; n_s + n_b)
\end{equation}

\noindent
allows us to construct dependences $n_s$ versus $n_b$ on given
value of Type II error $\beta \le \Delta$ (the probability that 
the observed number of events in planned experiment will be greater 
than
critical value $n_0$ if hypothesis $H_1$ is true) and given acceptance 
$1 - \alpha$ (the same probability if hypothesis $H_0$ is true).
If $\Delta = 2.85 \cdot 10^{-7}$ ($s \ge 5$, i.e. the value 
$n_0$ has $5\sigma$ deviation from average background $n_b$), 
the corresponding acceptance can be named {\it the probability of discovery}
and the dependence of $n_s$ versus $n_b$ - the $5\sigma$ discovery curve;
if $\Delta = 0.0014$ $(s \ge 3)$, the acceptance is {\it the probability of 
strong evidence}, and, if  $\Delta = 0.0228$ $(s \ge 2)$, the acceptance is
{\it the probability of weak evidence}. The case of weak evidence for
50\% acceptance ($s = 2$) is shown in Fig.1. 
The $5\sigma$ discovery, $3\sigma$ strong evidence, and $2\sigma$ weak
evidence curves for 90\% acceptance are presented in Fig.2.

\begin{figure}[htpb]

  \begin{center}
    \resizebox{6.2cm}{!}{\includegraphics{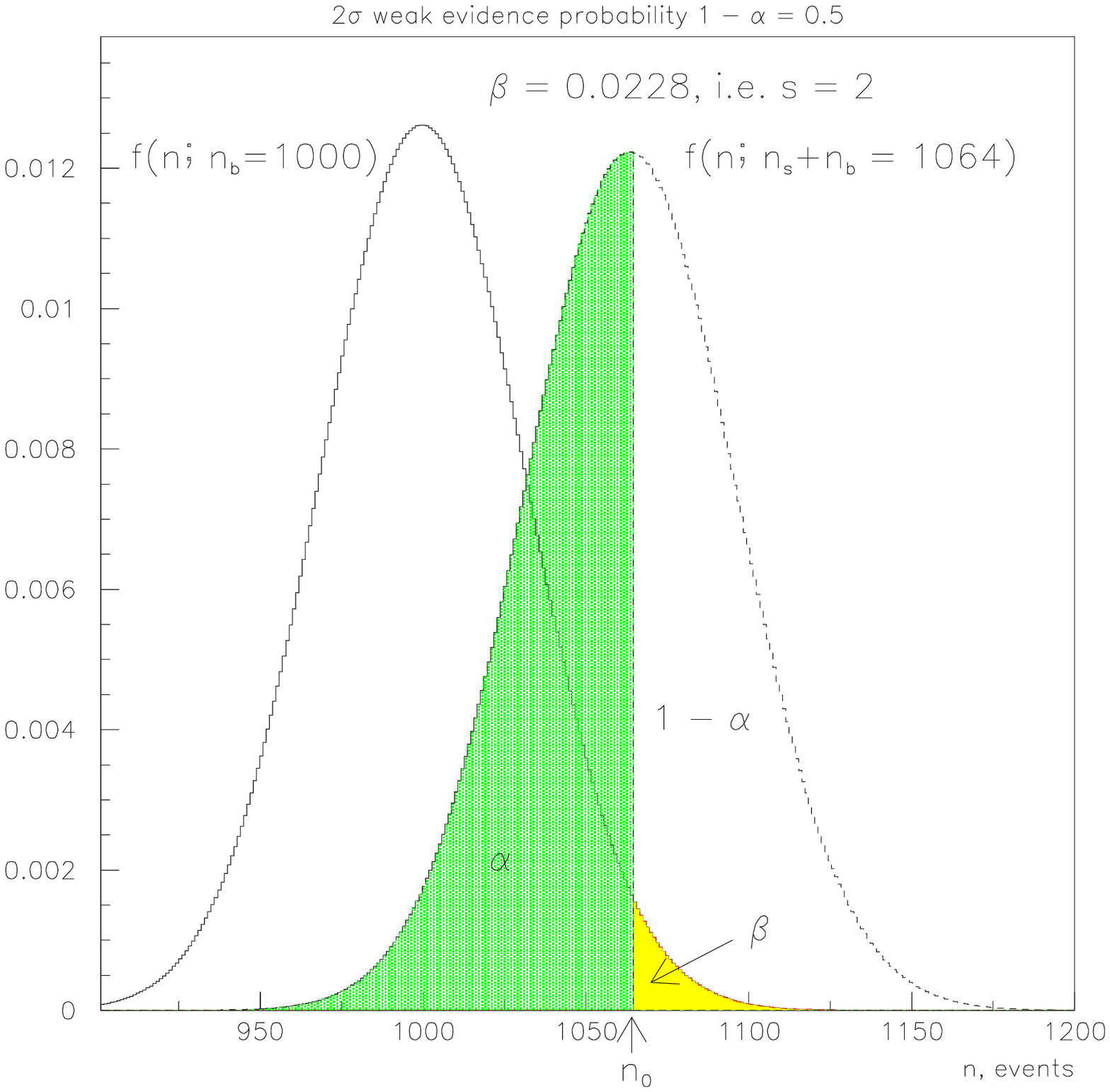}} 
\caption{The case $n_b \gg 1$. Poisson distributions with
parameters $\lambda = 1000$ (left) and $\lambda = 1064$ (right). Here 
$1 - \alpha = 0.5$ and $\beta = 0.02275$ (i.e. $s = 2$).}
    \label{fig:1} 
  \end{center}

  \begin{center}
    \resizebox{6.2cm}{!}{\includegraphics{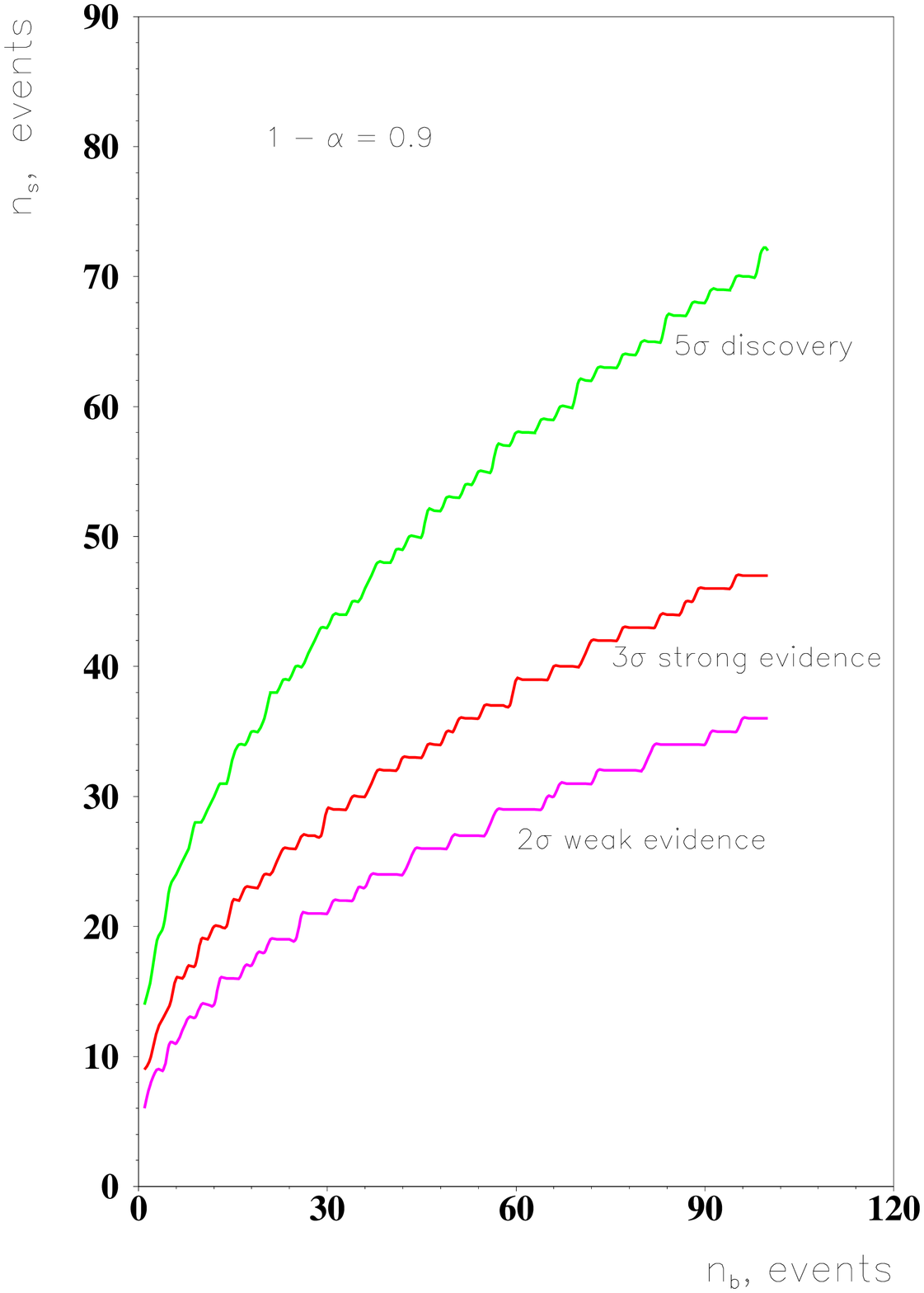}} 
\caption{Dependences $n_s$ versus $n_b$ for
$1 - \alpha = 0.9$ and for different values of $\beta$.}
    \label{fig:2} 
  \end{center}
\end{figure}

\section{Effects of one sided systematic errors on the discovery
potential}

We consider here forthcoming experiments to search for new physics. 
In this case we must take into account the systematic uncertainty 
which has theoretical origin without any statistical 
properties. For example, two loop corrections for most reactions at present 
are not known. In principle, it is ``reproducible
inaccuracy introduced by faulty technique''~\cite{Bevington}
and according to~\cite{Barlow} it contains the sense of
``incompetence''. If the predicted number of background events 
strongly exceeds the predicted number of signal events the discovery 
potential is the sensitive to this uncertainty. In this case 
we can only estimate the scale of influence of background uncertainty 
on the observability of signal, i.e. we can point the admissible level of 
uncertainty in theoretical calculations for given experiment proposal. 

Suppose uncertainty in the calculation of exact 
background cross section is determined by parameter $\delta$, i.e. the exact 
cross section lies in the interval $(\sigma_b, \sigma_b (1+\delta))$ 
and the exact value of the average number of background events 
lies in the interval 
$(n_b, n_b (1+\delta))$. Let us suppose $n_b \gg n_s$. 
As we know nothing about possible values of average number of 
background events, we consider the worst case~\cite{Bit1}. 
Taking into account 
formulae (3) and (4) we have the formulae

\begin{equation}
\beta =  \sum ^{\infty}_{n=n_0+1} f(n; n_b(1+\delta)) 
\leq \Delta
\end{equation} 

\begin{equation}
1 - \alpha = \sum ^{\infty}_{n = n_0 + 1}
f(n; n_b + n_s).
\end{equation}

\noindent
Formulae (5,6) realize the worst case when the background 
cross section $ \sigma_b(1 +\delta)$ 
is the maximal one, but we think that both the signal and the background 
cross sections are minimal.

The example of using these formulae
is shown in Fig.3.
We see  the sample of 200 (with, as expected, 100 background) 
events that will be 
enough to  reach 90\% probability of discovery with 25\% systematic 
uncertainty  of theoretical estimation of background.

\begin{figure}[htpb]
  \begin{center}
    \resizebox{6.2cm}{!}{\includegraphics{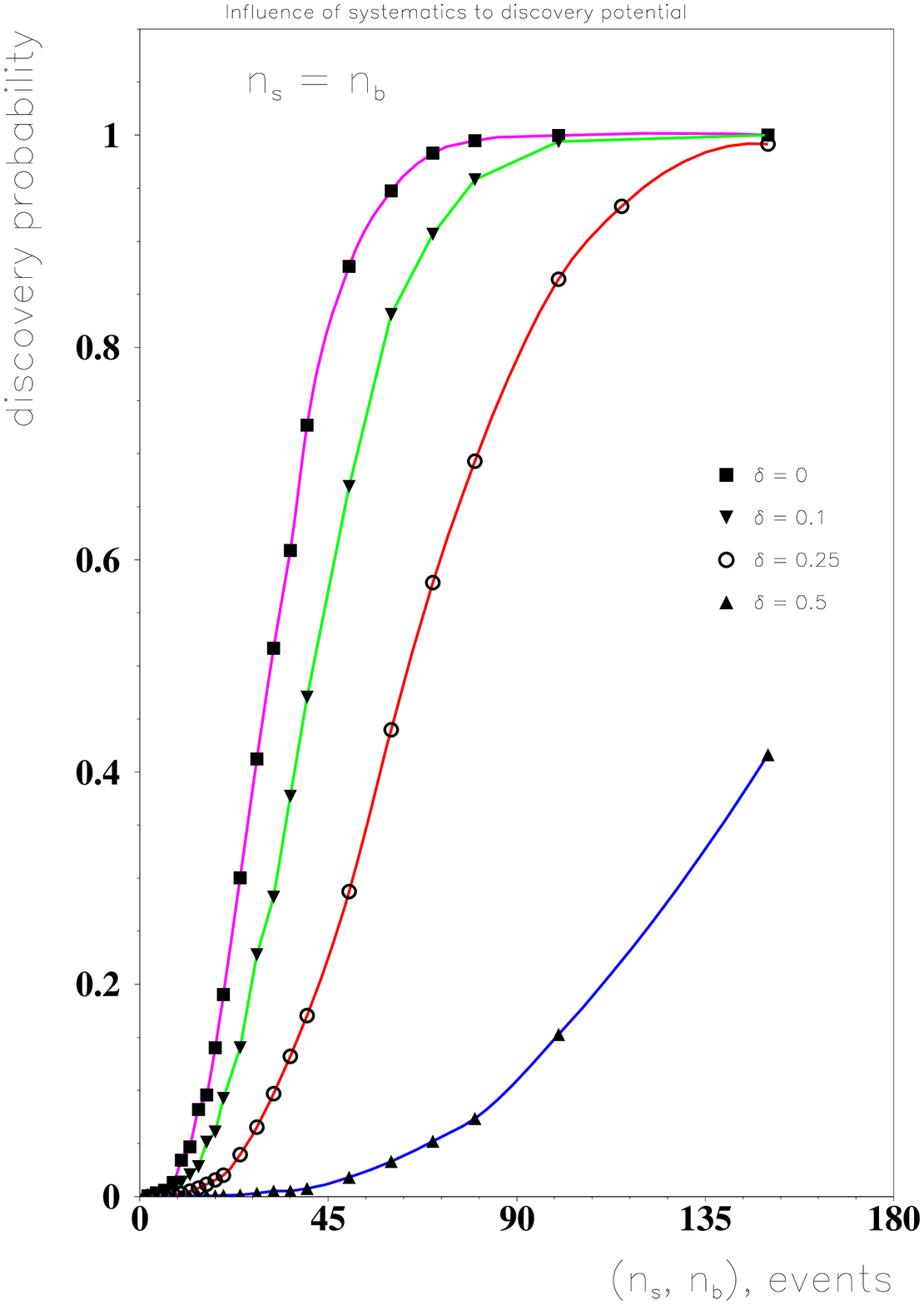}} 
\caption{Discovery probability versus $n_s$ for different values of
systematic uncertainty $\delta$ for the case $n_s=n_b$.
The curves are constructed under condition $\beta = 2.85 \cdot 10^{-7}$.}
    \label{fig:3} 
  \end{center}
\end{figure}

\section{An account of statistical uncertainty in the determination
of $n_s$ and $n_b$}

Usually, an experimentalist would extract the numbers $n_s$ and $n_b$ from
a Monte Carlo simulation of the planned experiment, which results in the
statistical errors.
If the probability of true value of parameter of Poisson distribution 
(the conditional probability) to be 
equal to any value of $\lambda \ge 0$ in the case when one observation 
$n_b = \hat n$ or $n_s + n_b = \hat n$ is known we have to take into account 
the statistical uncertainties in the determination of these values. 

Let us write down the density of Gamma distribution $\Gamma_{a, n + 1}$ as

\begin{equation}
g_n(a,\lambda) = \displaystyle 
\frac{a^{n+1}}{\Gamma(n+1)} e^{-a\lambda} \lambda^{n},   
\end{equation}

\noindent
where  $a$ is a scale parameter, $n + 1 > 0$ is a shape parameter, 
$\lambda > 0$ is a random variable, and $\Gamma(n+1) = n~!$ 
is a Gamma function. 

Let us set $a = 1$, then for each $n$ a continuous function

\begin{equation}
g_n(\lambda) = \displaystyle \frac{\lambda^n}{n!} e^{-\lambda},~ 
\lambda > 0,~n > -1  
\end{equation}

\noindent
is the density of Gamma distribution $\Gamma_{1, n + 1}$
with the scale parameter  $a = 1$ (see Fig.4). 
The mean, mode, and variance of 
this distribution are given by  $n+1,~n$, and $n+1$, respectively.

As it follows from the article~\cite{Why} 
and is clearly seen from the identity~\cite{Bit3} 
(Fig.5)

\begin{equation}
\displaystyle
\sum_{n = \hat n + 1}^{\infty}{f(n; \lambda_1)} +
\int_{\lambda_1}^{\lambda_2}{g_{\hat n}(\lambda) d\lambda} + 
\sum_{n = 0}^{\hat n}{f(n;\lambda_2)} = 1~,~~~i.e.
\end{equation}

$\displaystyle
\sum_{n = \hat n + 1}^{\infty}{\frac{\lambda_1^ne^{-\lambda_1}}{n!}} +
\int_{\lambda_1}^{\lambda_2}
{\frac{\lambda^{\hat n}e^{-\lambda}}{\hat n!}d\lambda}
+ \sum_{n = 0}^{\hat n}{\frac{\lambda_2^ne^{-\lambda_2}}{n!}} = 1~$ 

\noindent
for any $\lambda_1 \ge 0$ and 
$\lambda_2 \ge 0$,
the probability of true value of parameter of Poisson distribution
to be equal to the value of $\lambda$ in the case of one
observation $\hat n$ has probability density of 
Gamma distribution $\Gamma_{1,1+\hat n}$. The equation (9)
shows that we can mix Bayesian and frequentist probabilities in the given
approach.

\noindent

\begin{figure}[htpb]
  \begin{center}
        \resizebox{9.0cm}{!}{\includegraphics{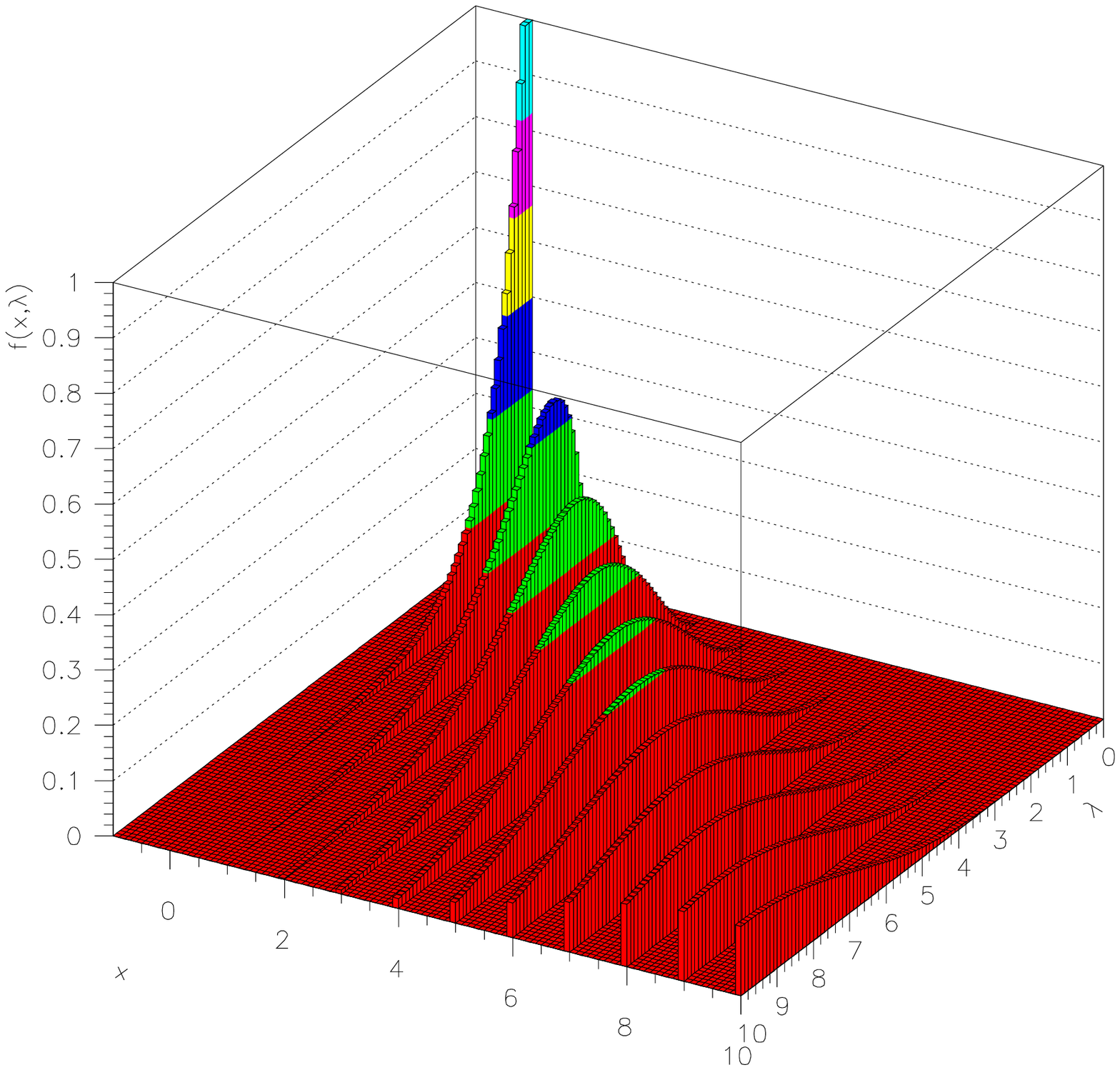}} 
\caption{The behaviour of the probability density of the true
 value of parameter $\lambda$ for the Poisson distribution 
 in case of $n$ observed events versus  $\lambda$ and  $n$. Here 
$f(n;\lambda)=g_n(\lambda)=\displaystyle \frac{\lambda^n}{n!}e^{-\lambda}$ 
 is both the Poisson distribution with the parameter $\lambda$ 
 along the axis  $n$ and the Gamma distribution with a shape 
 parameter $n+1$ and a scale parameter 1 along the axis $\lambda$.}
    \label{fig:4} 
  \end{center}
                
  \begin{center}
           \resizebox{9.0cm}{!}{\includegraphics{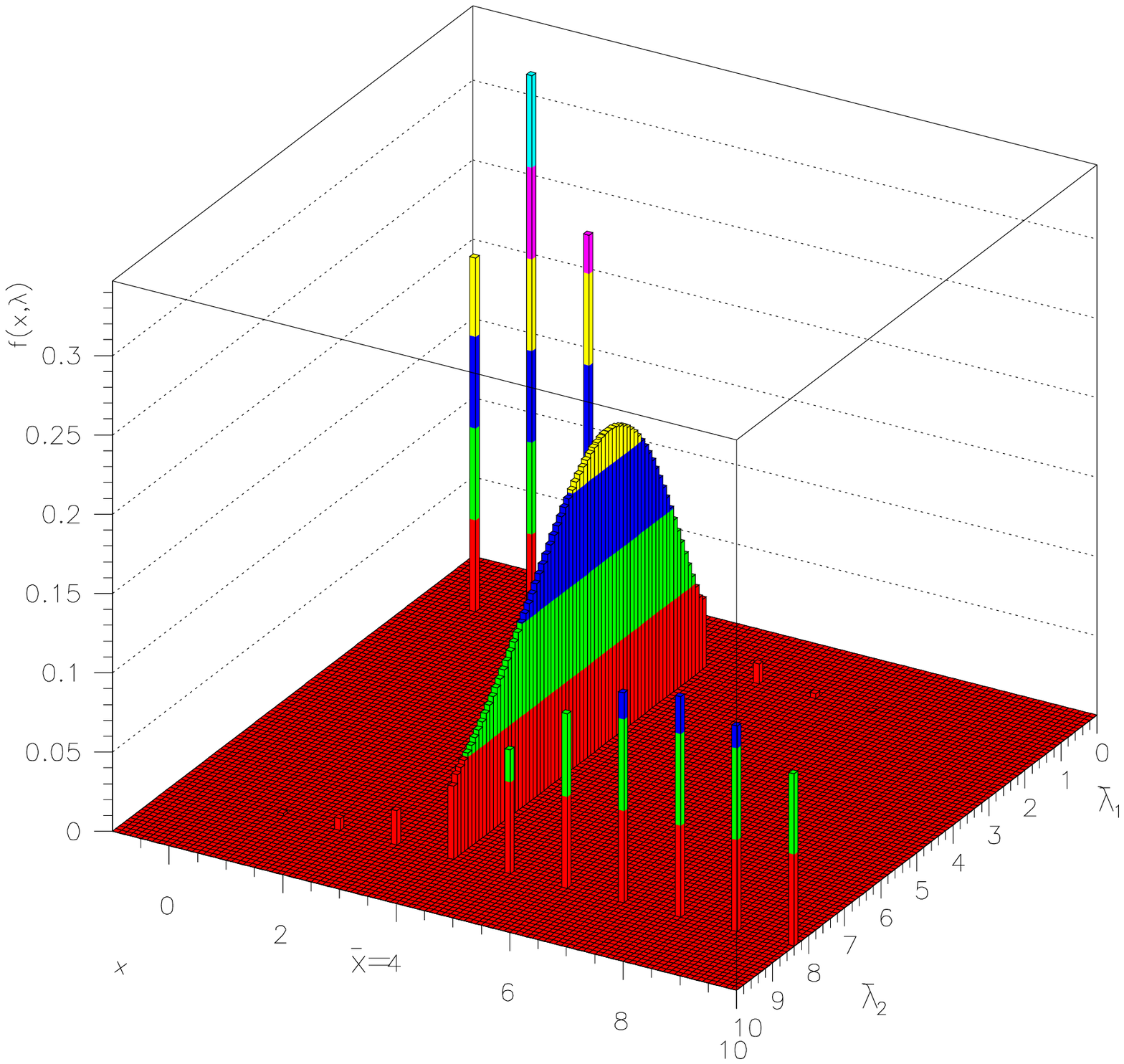}} 
\caption{The Poisson distributions $f(n,\lambda)$
for $\lambda$'s determined by the confidence limits 
$\hat \lambda_1 = 1.51$ and  $\hat \lambda_2 = 8.36$ 
in case of the observed number of events $\hat n = 4$ 
are shown. The probability density of Gamma distribution 
with a scale parameter  $a=1$ and a shape parameter  
$n+1=\hat n+1=5$ is shown within this confidence interval.}
    \label{fig:5} 
  \end{center}
\end{figure}

It allows to transform the probability distributions 
$f(n; n_s+n_b)$ and $f(n; n_b)$ accordingly
to calculate the probability of discovery~\cite{Bit4} 

\begin{equation}
\displaystyle 1 - \alpha = 1 -
\int_{0}^{\infty}{g_{n_s+n_b}(\lambda)\sum_{n=0}^{n_0}f(n;\lambda)d\lambda} 
= 1 - \sum_{n=0}^{n_0} \frac{C^n_{n_s+n_b+n}}{2^{n_s+n_b+n+1}}, 
\end{equation}

\noindent
where the critical value $n_0$ under the future 
hypotheses testing about the observability is chosen so that the Type II error

\begin{equation}
\displaystyle \beta =
\int_{0}^{\infty}{g_{n_b}(\lambda)\sum_{n=n_0+1}^{\infty}f(n; \lambda)d\lambda}
= \sum_{n=n_0+1}^{\infty} \frac{C^n_{n_b+n}}{2^{n_b+n+1}} 
\end{equation}

\noindent
could be less or equal to  $2.85 \cdot 10^{-7}$. 
Here $C^n_N$ is $\displaystyle \frac{N!}{n!(N-n)!}$.
Also we suppose that the Monte Carlo luminosity   
is exactly the same as the data luminosity later in the experiment.
The behaviour of discovery probability with and without account for
this uncertainty is shown in Fig.6. 

\begin{figure}[htpb]
  \begin{center}
    \resizebox{6.2cm}{!}{\includegraphics{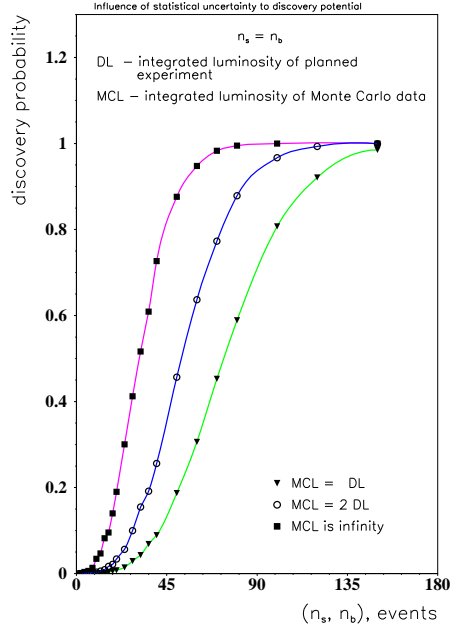}} 
\caption{Discovery probability versus $n_s$ with and without account for
statistical uncertainty in determination of $n_s$ and $n_b$. 
The case $n_s=n_b$.
The curves are constructed under condition $\beta = 2.85 \cdot 10^{-7}$.}
    \label{fig:6} 
  \end{center}
\end{figure}

The Poisson distributed random values 
have a property: if $\xi_i \sim Pois(\lambda_i),~i=1,2, \dots, m$ then 
$\displaystyle \sum^m_{i=1} \xi_i \sim 
\displaystyle Pois(\sum^m_{i=1} \lambda_i)$. 
It means that if we have $m$ observations 
$\hat n_1$, $\hat n_2$, $\dots$, $\hat n_m$ of the same random value 
$\xi \sim Pois(\lambda)$, 
we can consider these observations as one observation 
$\displaystyle \sum^m_{i=1} \hat n_i$  of the Poisson distributed random value 
with parameter $m \cdot \lambda$. According to eq.(9)
the probability of true value of parameter of this
Poisson distribution has probability density of 
Gamma distribution $\displaystyle \Gamma_{1,1+\sum_{i = 1}^{m}{\hat n_i}}$.
Using the scale parameter $m$ one can show that the probability of true value 
of parameter of Poisson distribution in the case of $m$ observations
of the random value $\xi \sim Pois(\lambda)$ 
has probability density of Gamma distribution 
$\displaystyle \Gamma_{m,1+\sum_{i = 1}^{m}{\hat n_i}}$, i.e. 
(see eq.7)

\begin{equation}
G(\sum{\hat n_i},m,\lambda) = g_{(\sum_{i = 1}^{m}{\hat n_i})}
(m,\lambda) = \displaystyle 
\frac{m^{(1+\sum_{i = 1}^{m}{\hat n_i})}}
{(\sum_{i = 1}^{m}{\hat n_i})!} e^{-m\lambda} 
\lambda^{(\sum_{i = 1}^{m}{\hat n_i})}. 
\end{equation}

Let us assume that the integrated luminosity of planned experiment is $L$
and the integrated luminosity of Monte Carlo data is $m \cdot L$. 
For instance, we can divide the Monte Carlo data into $m$ parts with
luminosity corresponding to the planned experiment. The result of Monte Carlo 
experiment in this case looks as set of $m$ pairs of numbers 
$(~(n_b)_i,~(n_b)_i+(n_s)_i~)$, where $(n_b)_i$ and $(n_s)_i$ are the numbers 
of background and signal events observed in each part of Monte Carlo data. 
Let us denote 
$\displaystyle N_b = \sum_{i=1}^m{(n_b)_i}$ and 
$\displaystyle N_{s+b} = \sum_{i=1}^m{((n_s)_i+(n_b)_i)}$.
Correspondingly~(see page 98,~\cite{Frodesen}),

\begin{equation}
\displaystyle \beta =
\int_{0}^{\infty}{G(N_b,m,\lambda)\sum_{n=n_0+1}^{\infty}
f(n; \lambda)d\lambda} =
\sum_{n=n_0+1}^{\infty} C^n_{N_b+n} \frac{m^{1+N_b}}{(m+1)^{1+N_b+n}}
\le \Delta, 
\end{equation}

\begin{equation}
\displaystyle 1 - \alpha = 1 -
\int_{0}^{\infty}{G(N_{b+s},m,\lambda)\sum_{n=0}^{n_0}
f(n;\lambda)d\lambda} = 1 - 
\sum_{n=0}^{n_0} C^n_{N_{s+b}+n} \frac{m^{1+N_{s+b}}}{(m+1)^{1+N_{s+b}+n}}. 
\end{equation}
 
\noindent
As a result, we have a generalized system of equations for the case
of different luminosity in planned data and Monte Carlo data.
The set of values 
$\displaystyle C^n_{N+n} \frac{m^{1+N}}{(m+1)^{N+n+1}},~n=0,1,\dots$ 
is a negative binomial (Pascal) distribution with real parameters 
$N+1$ and $\displaystyle \frac{1}{m+1}$, 
mean value $\displaystyle \frac{1+N}{m}$ and 
variance $\displaystyle \frac{(1+m)(1+N)}{m^2}$.

\section{Conclusions}

In this paper we have described  a method  to estimate  the discovery 
potential on new physics 
in planned experiments where only the average number of background $n_b$ and 
signal $n_s$ events is known. 
The ``effective significance'' $s$ of signal for given probability of
observation is discussed.
We also estimate the influence of systematic uncertainty related to non-exact 
knowledge of signal and background cross sections on the probability to 
discover new physics in planned experiments.
An account of such kind of systematics is very essential in the search 
for supersymmetry and leads to an essential decrease in the probability to 
discover new physics in future experiments. 
The texts of programs can be found in {\bf http://home.cern.ch/bityukov}.
A method for account of statistical uncertainties in determination 
of mean numbers of signal and background events is proposed.
Appendix A demonstrates the interrelation between Gamma- and Poisson 
distributions. The approach for estimation of 
exclusion limits on new physics is described in Appendix B. 

The author is grateful to N.V.~Krasnikov and V.F.~Obraztsov for 
the interest and useful comments, S.S.~Bityukov, 
Yu.P.~Gouz, V.V.~Smirnova, V.A.~Taperechkina
for fruitful discussions and E.A.~Medvedeva for help in preparing
the paper. The author thanks to referee of paper
for the constructive criticism. This work has been supported 
by grant CERN-INTAS 00-0440. 


\appendix

\section{The interrelation between gamma- and Poisson 
distributions}

The identity (9) (Fig.5)

$$
\displaystyle
\sum_{n = \hat n + 1}^{\infty}{f(n; \lambda_1)} +
\int_{\lambda_1}^{\lambda_2}{g_{\hat n}(\lambda) d\lambda} + 
\sum_{n = 0}^{\hat n}{f(n;\lambda_2)} = 1~,
$$

\noindent
can be easy generalized, as an 
example~\footnote{See, also, page 97 in ref.~\cite{Frodesen}, 
page 358 in ref.~\cite{Escoubes} and formula A7 in ref.~\cite{Silver}.}, 
to

$$
\displaystyle \sum_{n = k_m + 1}^{\infty}{f(n; \lambda_1)} + 
\sum^m_{i = 1}
[\int_{\lambda_i}^{\lambda_{i+1}} {g_{k_{m+1-i}}(\lambda) d\lambda} + 
\sum_{n = k_{m-i}+1}^{k_{m+1-i}}{f(n;\lambda_{i+1})}] 
$$

\bigskip

\begin{equation}
\displaystyle + f(k_0;\lambda_{m+1}) = 1 
\end{equation}

\noindent
for any real $\lambda_i \ge 0$, $i \in [1,m+1]$, integer 
$m>0$, $k_l > k_{l-1} \ge 0$, $l \in [1,m]$, $k_0 = 0$.

As a result of such type generalizations we have got

\begin{equation}
\displaystyle
\int_{\lambda_1}^{\lambda_2}{g_{m}(\lambda) d\lambda} + 
\sum_{i = n+1}^{m}{f(i;\lambda_2)} 
+ \int_{\lambda_2}^{\lambda_1}{g_{n}(\lambda) d\lambda} 
- \sum_{i = n+1}^{m}{f(i;\lambda_1)} = 0~,
\end{equation}

\noindent
i.e.

$\displaystyle\int_{\lambda_1}^{\lambda_2}
{\frac{\lambda^{m}e^{-\lambda}}{m!} d\lambda} + 
\sum_{i = n+1}^{m}{\frac{\lambda_2^i e^{-\lambda_2}}{i!}} 
+ \displaystyle \int_{\lambda_2}^{\lambda_1}
{\frac{\lambda^{n}e^{-\lambda}}{n!} d\lambda} 
- \sum_{i = n+1}^{m}{\frac{\lambda_1^i e^{-\lambda_1}}{i!}}  = 0~,$

\bigskip

\noindent
for any real $\lambda_1 \ge 0$, $\lambda_2 \ge 0$, and
integer $m > n \ge 0$.

\section{Exclusion limits~\cite{Bit1,Bit2}}

It is important to know the range in which a planned experiment can exclude 
presence of signal at given confidence level ($1 - \epsilon$).
It means that we will have uncertainty in future hypotheses testing
about non-observation of signal which equals to or less than $\epsilon$.
In refs.\cite{Hern,Taba} different methods to derive exclusion limits in 
prospective studies have been suggested. 

We propose to use the relative uncertainty

\begin{equation}
\tilde \kappa = \displaystyle \frac{\alpha+\beta}{2 - (\alpha+\beta)} 
\end{equation}
which will take place under hypotheses testing $H_0$ versus $H_1$.
It is {\it a probability of wrong decision}.
This probability $\tilde \kappa$ in case of applying 
{\it the equal-probability test}~\cite{Bit2} is a minimal relative value 
of the number of wrong 
decisions in the future hypotheses testing for Poisson distributions. It 
is the uncertainty in the observability of the new phenomenon. 
Note that in this case the probability of correct decision
$1 - \tilde \kappa$ (the relative number of correct decisions)
may be considered as a distance between two distributions
(the measure of distinguishability of two Poisson processes) 
in frequentist sense. This distance changes from zero up to unity
(as a result of the definition of equal-probability test).


\begin{thebibliography}{99}

\bibitem{PDG} Particle Data Group, D.E.~Groom et al., 
Eur.Phys.J. {\bf C15} (2000) 1 (Section 28).

\bibitem{Eadie} W.T.~Eadie, D.~Drijard, F.E.~James, M.~Roos, and B.~Sadoulet,
{\it Statistical Methods in Experimental Physics,} North Holland,
Amsterdam, 1971.

\bibitem{Bit1} S.I.~Bityukov and N.V.~Krasnikov,
{\it New physics discovery potential in future experiments,}
Modern Physics Letters {\bf A13}(1998)3235. 

\bibitem{Bit2} S.I.~Bityukov and N.V.~Krasnikov,
{\it On the observability of a signal above background},
Nucl.Instr.\&Meth. {\bf A452} (2000) 518;

S.I.~Bityukov and N.V.~Krasnikov,
{\it On observability of signal over background,}
Proceedings of Workshop on Confidence limits, 
YR, CERN-2000-005, 17-18 Jan., 2000, pp. 219-235,
eds. F.~James, L.~Lyons, Y.~Perrin. 

\bibitem{Sinervo} P.~Sinervo, 
{\it Signal significance in particle physics,} IPPP/02/39, DCPT/02/78,
Proceedings of International Conference ``Advanced Statistical 
Techniques in Particle Physics'', March 18-22, 2002, 
Durham, UK, p.64. http://www.ippp.dur.ac.uk/statistics/

\bibitem{Junk} T.~Junk,
{\it Confidence level computation for combining searches with
small statistics,} 
Nucl.Instr.\&Meth. {\bf A434} (1999) 435;

V.F.~Obraztsov,
{\it Confidence limits for processes with small statistics in
several subchannels and with measurement errors,}
Nucl.Instr.\&Meth. {\bf A316} (1992) 388;
Erratum, Nucl.Instr.\&Meth. {\bf A399} (1997) 500.

\bibitem{Frodesen} A.G.~Frodesen, O.~Skjeggestad, H.~T$\o$ft, 
{\it Probability and Statistics in Particle Physics,}
UNIVERSITETSFORLAGET, Bergen-Oslo-Troms$\o$, 1979, p.408.

\bibitem{Narsky} I.~Narsky, 
{\it Estimation of Upper Limits Using a Poisson Statistic,}
Nucl.Instrum.Meth. {\bf A450} (2000) 444.

\bibitem{Bit0}
S.I.~Bityukov, N.V.~Krasnikov,
{\it Uncertainties and Discovery Potential in Planned Experiments,}
IPPP/02/39, DCPT/02/78,
Proceedings of International Conference ``Advanced Statistical 
Techniques in Particle Physics'', March 18-22, 2002, 
Durham, UK, p.77. http://www.ippp.dur.ac.uk/statistics/;
e-Print: hep-ph/0204326, 2002. 

\bibitem{Bevington} R.~Bevington, 
{\it Data reduction and Analysis for the Physical Sciences},
McGraw Hill 1969.

\bibitem{Barlow} R.~Barlow, 
{\it Systematic errors: facts and fictions,} IPPP/02/39, DCPT/02/78,
Proceedings of International Conference ``Advanced Statistical 
Techniques in Particle Physics'', March 18-22, 2002, 
Durham, UK, p.134. http://www.ippp.dur.ac.uk/statistics/

\bibitem{Why} R.D.~Cousins {\it Why isn't every physicist a Bayesian ?}
{Am.J.Phys} {\bf 63} (1995) 398-410.

\bibitem{Bit3}
S.I.~Bityukov, N.V.~Krasnikov, V.A.~Taperechkina,
{\it Confidence intervals for Poisson distribution parameter,}
Preprint IFVE 2000-61, Protvino, 2000; also,
e-Print: hep-ex/0108020, 2001. 

\bibitem{Bit4} S.I.~Bityukov and N.V.~Krasnikov,
{\it Some problems of statistical analysis in experiment proposals,}
Proceedings of CHEP'01 International Conference on Computing 
in High Energy and Nuclear Physics, September 3-7, 2001, 
Beijing, P.R. China, Ed. H.S. Chen, Science Press,
Beijing New York, p.134. http://www.ihep.ac.cn/$\sim$chep01/

\bibitem{Escoubes} B.~Escoubes, S.~De~Unamuno and O.~Helene, {\it Experimental
signs pointing to a Bayesian instead of a classical approach
for experiments with a small number of events,}
Nucl.Instr.\&Meth. {\bf A257} (1987) 346.

\bibitem{Silver} D.~Silverman, {\it Joint Bayesian Treatment of
Poisson and Gaussian Experiments in a Chi-squared Statistic},
e-Print: physics/9808004, 1998 

\bibitem{Hern} J.J.Hernandez, S.Navas and P.Rebecchi,
{\it Estimating exclusion limits in prospective studies of searches,}
Nucl.Instr.$\&$Meth. A {\bf378}, 1996, p.301.

\bibitem{Taba} T.Tabarelli de Fatis and A.Tonazzo,
{\it Expectation values of exclusion limits in future experiments} (Comment),
Nucl.Instr.$\&$Meth. {\bf A403}, 1998, p.151.

\end{thebibliography}
\end{document}